\documentclass[aps,showpacs,amsmath,amssymb,pra,superscriptaddress,twocolumn,floatfix]{revtex4-1}

\usepackage{graphicx}
\usepackage{amsmath}
\usepackage{hyperref}
\usepackage{dcolumn}
\usepackage{bm}
\usepackage{color}

\usepackage{floatrow}

\begin{document}

\title{Local symmetries in one-dimensional quantum scattering}

\author{P.~A.~Kalozoumis}
\affiliation{Department of Physics, University of Athens, GR-15771 Athens, Greece}

\author{C.~Morfonios}
\affiliation{Zentrum f\"ur Optische Quantentechnologien, Universit\"{a}t Hamburg, Luruper Chaussee 149, 22761 Hamburg, Germany}

\author{F.~K.~Diakonos}
\affiliation{Department of Physics, University of Athens, GR-15771 Athens, Greece}

\author{P.~Schmelcher}
\affiliation{Zentrum f\"ur Optische Quantentechnologien, Universit\"{a}t Hamburg, Luruper Chaussee 149, 22761 Hamburg, Germany}
\affiliation{The Hamburg Centre for Ultrafast Imaging, Luruper Chaussee 149, 22761 Hamburg}

\date{\today}

\begin{abstract}
We introduce the concept of parity symmetry in restricted spatial domains -- local parity -- and explore its impact on the stationary transport properties of generic, one-dimensional aperiodic potentials of compact support. 
It is shown that, in each domain of local parity symmetry of the potential, there exists an invariant quantity in the form of a non-local current, in addition to the globally invariant probability current.
For symmetrically incoming states, both invariant currents vanish if weak commutation of the total local parity operator with the Hamiltonian is established, leading to local parity eigenstates.
For asymmetrically incoming states which resonate within locally symmetric potential units, the complete local parity symmetry of the probability density is shown to be necessary and sufficient for the occurrence of perfect transmission.
We connect the presence of local parity symmetries on different spatial scales to the occurrence of multiple perfectly transmitting resonances and propose a construction scheme for the design of resonant transparent aperiodic potentials. 
Our findings are illustrated through application to the analytically tractable case of piecewise constant potentials.
\end{abstract}

\pacs{03.65.-w, 73.40.Gk, 03.65.Xp, 73.63.Hs}
\maketitle

\section{Introduction \label{intro}}

\begin{figure*}[t!]
\floatbox[{\capbeside\thisfloatsetup{capbesideposition={right,top},capbesidewidth=5.5cm}}]{figure}[\FBwidth]
{\caption{(Color online) Local symmetries in 1D. (a) Schematic of a completely local parity symmetric potential, composed of three different arbitrary mirror-symmetric scatterers, labeled $A, B, C$, and intervening potential-free regions (gaps). The circular arcs above the scatterer array provide all possible spatial decompositions into local symmetry domains ${\mathcal D}_n$, of lengths $L_n/2$ and center positions $\alpha_n$, which completely cover the potential region (up to variations including part of the intervening gaps). Two selected decompositions are shown below the array (solid red and dashed green lines), demonstrating the presence of nested local symmetries within the same system. In a scattering setup, unit amplitude plane waves incident on the left (and on the right) are considered (see Sec.~\ref{scattering}), leading to transmitted $t$ (and $\tilde t$) and reflected $r$ (and $\tilde r$) amplitudes. (b) In one dimension, any translation of isolated constituents or parts of the potential can be reduced to combined overlapping local parity transformations: here, subsequent inversion through the centers of two subdomains, first $\mathcal D_1$, then $\mathcal D_2$.}\label{fig1}}
{\includegraphics*[width=11.5cm]{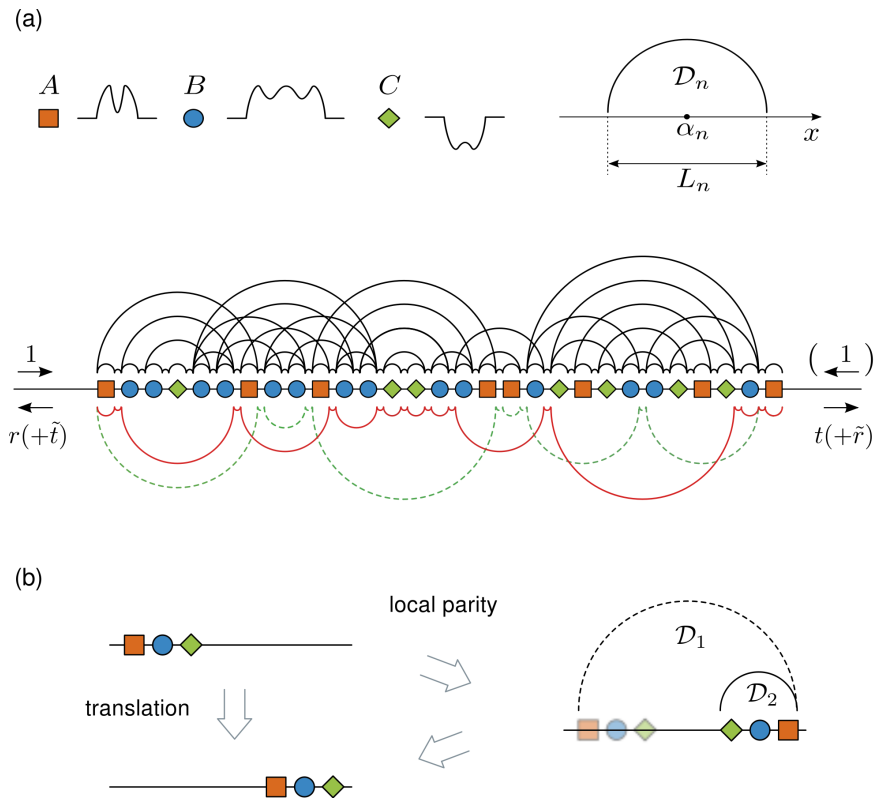}}
\vspace{.3cm}
\end{figure*}

Symmetries set the foundations upon which physical systems are treated, their ubiquitous presence on all scales dictating the form of the developed theory as well as experimental observation and analysis.
Global spatial symmetry, though, in most cases pertains exactly to structurally simple isolated systems and idealized models, while its absence in realistic situations often implies their statistical treatment~\cite{luna2009}.
To quantify situations of approximate symmetry, measures of symmetry~\cite{zabrodsky1992,pinsky2008} have been proposed, which reflect the degree to which it is fulfilled under specific operations.
In the generic case, symmetry of a system under spatial transformations is globally broken, but retained on a local scale~\cite{echeverria2011}, such that the associated invariance of physical properties affects its behavior beyond a mean description~\cite{Coupier2005}.
If a system can be completely covered by spatial domains where its structure exhibits such local symmetry, it can be regarded as {\it completely locally symmetric}.
Since these domains can be of variable extent and at different locations within a single system, there is, in general, a multitude of possible local symmetry decompositions with different symmetry scales and axes (see Fig.~\ref{fig1}(a)).
Local symmetries can be present by design, as in electronic transport devices ~\cite{Ferry1997} and dielectric multilayers in nanophotonics \cite{Macia2006,Macia2012,Zhukovsky2010,Huang2001, Peng2002,Peng2004,Hsueh2011}, but also naturally inherent in structurally complex systems, like large molecules \cite{Grzeskowiak1993,Pascal2001,Chen2012}, quasi-crystals \cite{Shechtman1984,Levine1984,Widom1989,Lifshitz1996}, or even disordered matter~\cite{Wochner2009}.

In spite of the intensive exploration of such systems, their properties have, to a large extent, been related only to global symmetry characteristics.
Although classification and study of local {\it structural} features have been carried out, with consequences for spectral and localization properties \cite{Macia2012}, little attention has been paid to the impact of explicit local {\it symmetries} -- even if they are very obviously present.
In particular, perfect transmission \cite{Peng2002,Peng2004,Huang2001,Zhukovsky2010,Hsueh2011} through a scattering potential is commonly connected to its global symmetry, whereas the role of local symmetries for resonant states is disregarded.
Nevertheless, it is readily seen schematically in Fig.~\ref{fig1}(a) that global symmetry is a (trivial) special case among -- and not a necessity for -- a plethora of possible local symmetry decompositions of a potential.
The question thus arises, whether and how these local symmetries affect its (resonant) scattering properties, in the {\it absence} of global symmetry.
In general, the need for a rigorous theoretical treatment which addresses local regularities, but incorporates the global composite structure, becomes evident.

In this work we take a step in this direction, investigating the impact of local symmetries in the transport properties of globally non-symmetric systems.
Considering scattering in one dimension, any euclidean transformation can be reduced to combinations of coordinate inversions within subintervals of configuration space (see Fig.~\ref{fig1}(b)), which we refer to as {\it local parity} (LP) transformations.
Utilizing completely LP symmetric potentials, we show how LP is related to the transport properties of stationary scattering states with symmetric or asymmetric asymptotic conditions (SAC or AAC, see Fig.~\ref{fig1}(a)): under conservation of LP, the total or reflected probability current vanishes, respectively.

Specifically, it is proven that (i) states with definite LP, which can arise only with waves incident symmetrically on both sides of the potential, have zero probability current, and, more importantly, that (ii) with one incident wave, the scattering state exhibits a perfect transmission resonance (PTR) which resonates within locally symmetric potential units, if and only if its probability density is completely LP symmetric.
These PTRs are classified with respect to the reducibility of the density profile into local symmetries at lower scale, i.e. smaller LP symmetric domains, following those of the potential.
In this sense, it is shown that for any such PTR there {\it exists} an irreducible decomposition of the density profile into locally resonating units.
We stress that the introduced concepts and the derived results hold for {\it arbitrary} one-dimensional (1D) LP symmetric potentials of compact support, thereby enabling the development of a construction principle for PTRs at desirable energies in globally non-symmetric setups. 
As an illustrative example we consider the analytically tractable case of piecewise constant (PWC) potentials.

The paper is organized as follows: 
In Sec.~\ref{local_parity} the LP operators are defined and analyzed. 
In Sec.~\ref{scattering}, LP is applied to scattering in arbitrary 1D potentials, and its relation to zero-current states and PTRs is developed. 
In Sec.~\ref{construction} we propose a general construction principle for setups with PTRs at prescribed energies, which we use to illustrate our findings in representative examples of PWC potentials.
Section \ref{conclusion} concludes the paper.

\section{Local parity \label{local_parity}}

Let us first introduce the LP operation, which performs the usual parity transform in a finite subdomain $\mathcal{D}$ of configuration space (of the $x$-axis in 1D) and acts in the remaining part, up to a sign, as the identity operator, thereby maintaining spectral equivalence to global parity. 
The single LP operator $\hat{\varPi}^\mathcal{D}_s \equiv\hat{\varPi}_{s}^{(L,\alpha)}$, with $s = \pm 1$, is parametrized by the location of its inversion point $\alpha$ and the size $L$ of $\mathcal{D}$, in terms of which it is defined by its action on an arbitrary state $|\psi(x)\rangle$ as 
\begin{eqnarray}
\hat{\varPi}_{s}^{\mathcal{D}} |\psi(x)\rangle ~ = ~&& {\varTheta} \left(\frac{L}{2} -\vert x -\alpha \vert \right)|\psi(2\alpha-x)\rangle  \nonumber \\ 
+ ~s~ && {\varTheta} \left(\vert x -\alpha \vert - \frac{L}{2} \right)|\psi(x)\rangle, ~~~~ s = \pm 1~~
\label{lp_def}
\end{eqnarray}
where $\varTheta$ denotes the Heaviside step function.
That is, in addition to inverting the argument of $|\psi(x)\rangle$ within $\mathcal D$, $\hat{\varPi}_{-}^{\mathcal{D}}$ changes its sign outside $\mathcal D$, while $\hat{\varPi}_{+}^{\mathcal{D}}$ retains it.

\begin{figure}[b!]
\includegraphics*[width=\columnwidth]{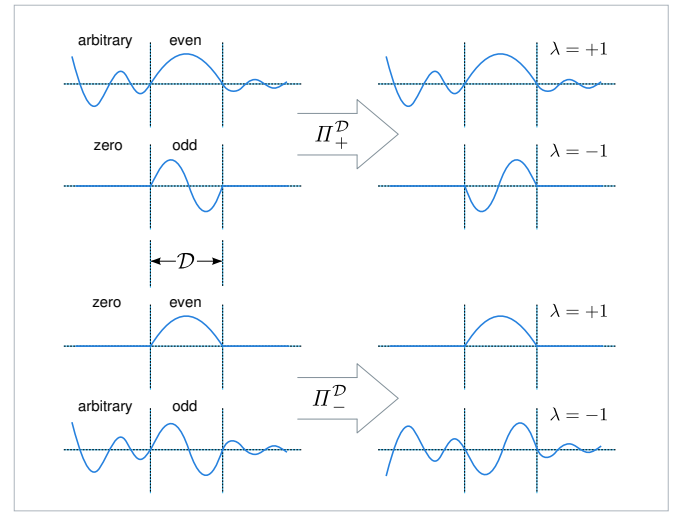}
\caption{\label{fig2} Local parity eigenstates, (their real parts) schematically shown for a single subdomain $\mathcal{D}$. Eigenstates of $\hat{\varPi}_{+}^{\mathcal{D}}$ ($\hat{\varPi}_{-}^{\mathcal{D}}$) with eigenvalue $\lambda = +1$ ($-1$) are even (odd) within $\mathcal{D}$ and arbitrary outside. Eigenstates of $\hat{\varPi}_{+}^{\mathcal{D}}$ ($\hat{\varPi}_{-}^{\mathcal{D}}$) with eigenvalue $\lambda = -1$ ($+1$) are odd (even) within $\mathcal{D}$ and zero outside.}
\end{figure}

The LP operator can thus be regarded as a generalization of the global parity operator, to which it reduces for $L \to \infty$.
Since $(\hat{\varPi}_{s}^{\mathcal{D}})^2 = {\hat 1}$, either one of the two operators has two eigenvalues, $\lambda_s = \pm 1$.
The corresponding sets of eigenstates of the $\hat{\varPi}_{s}^{\mathcal{D}}$ each have two parts, one odd and one even, as shown schematically in Fig.~\ref{fig2}.
The even (odd) eigenstates of $\hat{\varPi}_{-}^{\mathcal{D}}$ ($\hat{\varPi}_{+}^{\mathcal{D}}$) necessarily vanish outside $\mathcal D$, thus corresponding to isolated bound states with {\it global} definite parity.
In contrast, even (odd) eigenstates of $\hat{\varPi}_{+}^{\mathcal{D}}$ ($\hat{\varPi}_{-}^{\mathcal{D}}$) are arbitrary outside ${\mathcal D}$, and can therefore non-trivially possess {\it local} parity; in particular, they are relevant for scattering, since they allow for open boundary conditions.

Two different LP transforms $\hat{\varPi}^{\mathcal{D}_1}_{s_1}$, $\hat{\varPi}^{\mathcal{D}_2}_{s_2}$ commute if the associated domains do not overlap, $\mathcal{D}_1 \cap \mathcal{D}_2 = \oslash$.
Subsequent application of $N = N_+ + N_-$ non-overlapping single LP transforms, where $N_\pm$ is the number of acting $\hat{\varPi}^{\mathcal{D}_n}_\pm$ operators, thus corresponds to a {\it total} LP operator
\begin{equation}
\hat{\varPi} = \prod_{n=1}^{N} \hat{\varPi}_{s_n}^{\mathcal{D}_n}, ~~~ s_n = \pm 1
\label{lp_total}
\end{equation}
having again two eigenvalues $\lambda = \prod_{n=1}^{N}\lambda_{s_n} = \pm 1$, each of which has degeneracy $2^{N-1}$ (being the sum of odd binomials).
As a consequence of the properties of the LP eigenstates, seen in Fig.~\ref{fig2}, an eigenstate of $\hat{\varPi}$ can be non-vanishing only in a {\it single} subdomain $\mathcal{D}_n$ if it is an eigenstate of $\hat{\varPi}_{s_n = \pm 1}^{\mathcal{D}_n}$ with {\it opposite} eigenvalue $\lambda_n = \mp 1$.
Therefore, an eigenstate of $\hat{\varPi}$ (with eigenvalue $\lambda = (-1)^{N_-}$) is non-vanishing in multiple subdomains, and thereby relevant for scattering, only if it is a simultaneous eigenstate of each $\hat{\varPi}^{\mathcal{D}_n}_{s_n}$ with eigenvalue $\lambda_n = s_n$.

As we will see in the following, the link between local symmetry and transport properties is provided by the commutation of the Hamiltonian $\hat{H} = \frac{1}{2}\hat{k}^2 + V(\hat{x})$ (using units $\hbar = m = 1$, with an arbitrary energy unit $\epsilon$) with $ \hat{\varPi}$, and thereby with each $\hat{\varPi}_{s_n}^{\mathcal{D}_n}$.
Consider a completely LP symmetric potential $V(x)$, i.e., one which is symmetric about $\alpha_n$ within every subdomain $\mathcal{D}_n = [\alpha_n - \frac{L_n}{2}, \alpha_n + \frac{L_n}{2}]$, 
\begin{eqnarray}
V(x) = V(2\alpha_n - x) ~~~ \forall x \in {\mathcal D}_n, ~~~ n=1,2,...,N,
\label{pot_symmetry}
\end{eqnarray}
for a given spatial decomposition into $N$ subdomains.
The action of the two commutators associated with the $n$-th subdomain then reads
\begin{widetext}
\begin{equation}
[\hat{H},\hat{\varPi}_{s_n}^{\mathcal{D}_n}]~ |\psi(x)\rangle ~=~ \frac{1}{2}~\Delta'(x) \left\{ |\psi(2 \alpha_n - x)\rangle - s_n |\psi(x)\rangle \right\} - \Delta(x) \left\{ |\psi(2 \alpha_n - x)\rangle'  + s_n |\psi(x)\rangle' \right\}, ~~ s_n = \pm 1
\label{lp_comm}
\end{equation}
\end{widetext}
where $\Delta(x) = \delta (x-\alpha_n-\frac{L_n}{2}) - \delta(x-\alpha_n+\frac{L_n}{2})$ is a sum of boundary Dirac $\delta$-functions and the prime denotes differentiation with respect to $x$.
Note that, whereas $[\hat{\varPi}_{s_n}^{\mathcal{D}_n},\hat{V}]|\psi(x)\rangle = [V(x) -  V(2\alpha_n - x) ]|\psi(2\alpha_n - x)\rangle = 0$ for $x \in {\mathcal D}_n$ by assumption, the kinetic term in $\hat{H}$ leads to non-vanishing boundary terms.
The commutation is then manifest in a weak sense, that is, subsequent action of $\hat{H}$ and each $\hat{\varPi}_{s_n}^{\mathcal{D}_n}$ on $\psi(x)$ is independent of their order, only if the right hand side of Eq.(\ref{lp_comm}) vanishes. 
Then, $\hat{H}$ commutes weakly also with the total LP operator $\hat{\varPi}$, since 
\begin{eqnarray}
[\hat{H},\hat{\varPi}]|\psi(x)\rangle &=& [\hat{H},\prod_{n=1}^{N-1} \hat{\varPi}_{s_n}^{\mathcal{D}_n}]\hat{\varPi}_{s_N}^{\mathcal{D}_N}|\psi(x)\rangle \nonumber \\ &=& \lambda_{s_N}[\hat{H},\prod_{n=1}^{N-2} \hat{\varPi}_{s_n}^{\mathcal{D}_n}]\hat{\varPi}_{s_{N-1}}^{\mathcal{D}_{N-1}}|\psi(x)\rangle \nonumber \\ &=& \left(\prod_{n=2}^{N}\lambda_{s_n}\right)[\hat{H},\hat{\varPi}_{s_1}^{\mathcal{D}_1}]|\psi(x)\rangle = 0.
\end{eqnarray}

Because of the local symmetry of the potential, Eq.~(\ref{pot_symmetry}), the commutation relation in Eq.~(\ref{lp_comm}) must hold inwards for any $L \leqslant L_n$ within ${\mathcal{D}_n}$. 
We thus conclude that $|\psi(x)\rangle$ is a common eigenstate of $\hat{H}$ and $\hat{\varPi}$, and therefore LP definite, if its wave function fulfills the $N$ conditions
\begin{eqnarray}
\psi(x) &=& s_n \psi(2\alpha_n - x) ~,~~ x \in {\mathcal D}_n, ~~ n = 1,2,...,N.
\label{lp_cond}
\end{eqnarray}
In polar representation, $\psi(x) = \sqrt{\rho(x)}e^{i\varphi(x)}$, this yields the $2N$ conditions
\begin{subequations}
 \begin{align}
  \rho(x) &= \rho(2\alpha_n - x) \label{lp_cond_rho} \\ \varphi(x) &= \varphi(2\alpha_n - x) + \frac{(1 - s_n)\pi}{2} \label{lp_cond_phi}
 \end{align}
\end{subequations}
for the probability density $\rho(x)$ and for the phase $\varphi(x)$ defined up to mod$(2\pi)$. 
Violating one of these two conditions implies a breaking of LP symmetry.

Note here that spectral equivalence of the LP operators with global parity (i.e., having global eigenvalues $\lambda = \pm 1$) is ensured at the expense of strong (state independent) commutation between $\hat{H}$ and $\hat{\varPi}$.
In other words, the weakness of the commutation in Eq.~(\ref{lp_comm}) is the price to pay for a {\it local} parity operator with {\it global} (i.e., defined over all $x \in \mathbb R$) action and eigenvalues.

The wave function $\psi_k(x)$ of a stationary state with energy $E = k^2/2$ is a solution of the time-independent Schr\"odinger equation
\begin{equation}
 \frac{1}{2} \psi_k''(x) + \gamma^2(x)\psi_k(x) = 0, ~~ \gamma^2(x) = \frac{k^2}{2}  - V(x),
\label{SE_psi}
\end{equation}
which, in the polar representation, is transformed to a non-linear equation for the modulus $u_k(x) = |\psi_k(x)| = \sqrt{\rho_k(x)}$,
\begin{equation}
 \frac{1}{2} u_k''(x) - \frac{j_k^2}{2u_k^3(x)} + \gamma^2(x)u_k(x) = 0, 
\label{SE_mod}
\end{equation}
where
\begin{equation}
 j_k = u_k^2(x) \varphi_k'(x) = \rho_k(x) \varphi_k'(x) = {\rm const}
\label{current}
\end{equation}
is the spatially invariant ($x$-independent) probability current.
Note that, if $u_k(x)$ has nodes, $j_k$ necessarily vanishes, so that no divergence occurs in Eq.~\ref{SE_mod}.

While current invariance holds for any 1D potential, the special case of a completely LP symmetric potential, Eq.~(\ref{pot_symmetry}), yields yet another {\it locally} invariant quantity, in the form of a complex non-local current:
Restricting Eq.~(\ref{SE_psi}) to the subdomain ${\mathcal D}_n$, multiplying by $\psi_k(2\alpha_n - x)$ and subtracting the LP transformed ($x \to 2\alpha_n - x$) result, leads to
\begin{eqnarray}
 &\psi_k(x)&\psi_k'(2\alpha_n - x) \nonumber \\ + ~&\psi_k'(x)&\psi_k(2\alpha_n - x) \equiv  q_{k,n} = {\rm const},
\label{q}  
\end{eqnarray}
provided that $V(x) = V(2\alpha_n - x)$ for $x \in {\mathcal D}_n$.
In the case of a completely LP symmetric potential, there are $N$ (generally different) such constants $q_{k,n}$, with values depending on the decomposition into $N$ subdomains.

If the current $j_k$ vanishes, Eq.~(\ref{SE_mod}) becomes identical to Eq.~(\ref{SE_psi}), although now for the real field $u_k(x)$. Using the same procedure as for Eq.~(\ref{q}) we find a (real) invariant quantity for the wave function modulus,
\begin{eqnarray}
 &u_k(x)&u_k'(2\alpha_n - x) \nonumber \\ + ~&u_k'(x)&u_k(2\alpha_n - x) \equiv  {\tilde q}_{k,n} = {\rm const}.
\label{qtilde}
\end{eqnarray}

The non-local quantity $q_{k,n}$ (and its counterpart ${\tilde q}_{k,n}$ for zero current) expresses the manifestation of LP symmetry in the state of the system: It remains invariant under the LP transformations of $\hat\varPi$ for {\it any} wave function $\psi_k(x)$,  while it vanishes for LP eigenstates.

We now proceed to relate even further the spatially invariant (zero or non-zero) quantities $j_k$ and $q_{k,n}$ to the LP of energy eigenstates.
To this aim, we substitute the polar form of $\psi_k(x)$ into Eq.~(\ref{q}) and separate real and imaginary parts, which yields the following equations:
\begin{subequations}
\begin{align}
&u_k(x) u_k'(2\alpha_n-x)+u_k'(x) u_k(2\alpha_n-x) \nonumber \\ 
&~~~~~~~= |q_{k,n}|\cos(\vartheta_{k,n}-\varphi_k(x)-\varphi_k(2\alpha_n-x)), \label{class1} \\
&j_{k}\left(\frac{u_{k}(2\alpha_n-x)}{u_{k}(x)}+\frac{u_{k}(x)}{u_{k}(2\alpha_n-x)}\right) \nonumber \\ 
&~~~~~~~= |q_{k,n}|\sin(\vartheta_{k,n}-\varphi_k(x)-\varphi_k(2\alpha_n-x)), \label{class2}
\end{align}
\end{subequations}
where $q_{k,n} \equiv |q_{k,n}| e^{i\vartheta_{k,n}}$.

The two cases (i) $j_k=0$ and (ii) $j_k \neq 0$ are now distinguished, which classify the possible scattering states with respect to their LP properties depending on the invariant quantities $q_{k,n}$, as follows.

(i) $j_k=0$:
Then $q_{k,n} = \tilde q_{k,n} = 0$ or $\vartheta_{k,n}-\varphi_k(x)-\varphi_k(2\alpha_n-x)=m \pi ~(m\in\mathbb{Z})$, as seen from Eqs.~(\ref{class2}) and (\ref{qtilde}).
If the former condition holds for all subdomains $\mathcal D_n$, then $\psi_{k}(x)$ has definite total LP , i.e., it is an LP eigenstate of $\hat{\varPi}$, due to Eqs.~(\ref{q}) and (\ref{lp_cond}).
Otherwise, $\tilde q_{k,n} = \pm|q_{k,n}|$ from Eqs.~(\ref{class1}) and (\ref{qtilde}), and $\psi_{k}(x)$ is simply a state with zero probability current but without definite total LP.

(ii) $j_k \neq 0$: Then $u_k(x) \neq 0 ~\forall x \in \mathcal D_n$ in Eq.~(\ref{class2}) (which excludes totally reflected states under AAC, see next section), so that also $q_{k,n} \neq 0$.
Now, if additionally $\vartheta_{k,n}-\varphi_k(x)-\varphi_k(2\alpha_n-x)= (m + \frac{1}{2}) \pi ~(m\in\mathbb{Z})$ in each $\mathcal D_n$, then $u_k(x) u_k'(2\alpha_n-x) = -u_k'(x) u_k(2\alpha_n-x)$ from Eq.~(\ref{class1}), corresponding to a completely LP symmetric probability density $\rho_k = u_k^2$.
This is the case of a PTR, as will be shown in the next section.

\section{Scattering \label{scattering}}

We now employ LP to study 1D scattering in a completely LP symmetric potential of the form
\begin{equation}
 V(x) = \sum_{n=1}^{N_\mathcal{S}} V_n(x) ~{\varTheta}(L_n/2-|x-\alpha_n| )
\label{potential}
\end{equation}
describing an array of $N_\mathcal{S}$ non-overlapping scatterers ${\mathcal S}_n$ of widths $L_n$, centered at and symmetric about $x = \alpha_n$.
The scatterer potentials $V_n$ thus obey Eq.~(\ref{pot_symmetry}) with $N = N_\mathcal{S}$, but are otherwise {\it arbitrary} bounded functions ($|V_n(x)|<\infty$).
Such a potential is shown schematically in Fig.~\ref{fig1}(a) for three different types of scatterers.
Obviously, $V(x)$ incorporates the case of a globally symmetric potential if the scatterers are distributed symmetrically with respect to the array center.
Also, the potential-free gaps are not essential, meaning that $V(x)$ could as well represent a {\it single} continuous LP symmetric scatterer if the gap lengths are set to zero.

The stationary scattering solution $\psi_k(x)$ of Eq.~(\ref{SE_psi}), and thereby $u_k(x)$ of Eq.~(\ref{SE_mod}), are uniquely determined by the asymptotic conditions at $x \to \pm\infty$, which coincide with boundary conditions at the ends $x = \alpha_N + L_N/2, \alpha_1 - L_1/2$ of the potential since it is 1D and explicitly 'unbiased' (of compact support). 
Symmetric and asymmetric asymptotic conditions are imposed with plane waves $e^{\pm ikx} = \langle x|\pm \rangle$ incident on the left and right, or only on the left, of the scattering region, respectively (where $k > 0$).
The corresponding asymptotic ingoing amplitudes $\bigl(\begin{smallmatrix} 1 \\ 1 \end{smallmatrix}\bigr)$ or $\bigl(\begin{smallmatrix} 1 \\ 0 \end{smallmatrix}\bigr)$ in $|\pm \rangle$-space are scattered off the potential into the outgoing ones by the $S$-matrix
\begin{equation}
 S = \left( \begin{array}{cc} r & t \\ t & \tilde{r} \end{array} \right) \equiv e^{i\zeta}\left( \begin{array}{cc} \sqrt{R}e^{i\eta} & \sqrt{T} \\ \sqrt{T} & -\sqrt{R}e^{-i\eta} \end{array} \right)
\label{s_matrix}
\end{equation}
which is unitary by momentum (probability current) conservation and symmetric due to time-reversal invariance. 
$T = |t|^2 = 1 - |r|^2 = 1 -R$ is the transmission probability, which is measurable for a setup under AAC.

Both for SAC and AAC, the role of the current in Eq.~(\ref{current}) is decisive for the present analysis, because it relates the scattering properties of the system to symmetries in $\psi_k(x)$:
For $j_k \neq 0$, any (local) symmetry in $\rho_k(x)$ implies the same symmetry in $\varphi_k'(x)$.

\subsection{LP eigenstates with zero probability current \label{zcs}}

If $\psi_k(x)$ is an LP definite eigenfunction of $\hat{\varPi}$, then it should have a locally antisymmetric phase slope $\varphi_k'(x)$, as implied by Eq.~(\ref{lp_cond_phi}).
This is incompatible with Eq.~(\ref{current}) for locally symmetric $\rho_k(x)$, unless $\varphi_k'(x)=0$.
Therefore, $\varphi_k(x)$ must be constant where $\rho_k(x) \neq 0$ with $\pm\pi$-jumps at nodes of $\rho_k(x)$, so that $\psi_k(x)$ maintains its LP.
The state then carries zero current.

Considering scattering off the potential $V(x)$, an LP eigenstate can only be realized under SAC, since AAC break LP symmetry explicitly, as will be shown.
To confirm the impact of definite LP on the $S$-matrix, we use SAC with wave function $\psi^<_k(x) = e^{ikx} + (t+r)e^{-ikx}$ on the left and $\psi^>_k(x) = e^{-ikx} + (t+\tilde{r})e^{ikx}$ on the right of the array.
LP conservation is now imposed in all subdomains $\mathcal{D}_n = [\alpha_n - \frac{L_n}{2}, \alpha_n + \frac{L_n}{2}] ~ (n = 1,2,...,N_{\mathcal{S}})$ of the scatterers, and in the subdomains $\mathcal{\bar D}_{\bar n} = [\alpha_{\bar n} + \frac{L_{\bar n}}{2}, \alpha_{\bar n + 1} - \frac{L_{\bar n + 1}}{2}]~ (\bar n = 1,2,...,N_{\mathcal{S}}-1)$ between them, through condition (\ref{lp_cond}). 
The partial eigenvalues are $\lambda_n = s_n (\lambda_{\bar n} = s_{\bar n})$ in each $\mathcal{D}_n (\mathcal{\bar D}_{\bar n})$, so that the state is not isolated within a single subdomain and has total eigenvalue $\lambda = (-1)^{N_-}$, where $N_-$ is the number of applied $s_{n} = -1$ and $s_{\bar n} = -1$ operations, as discussed above.

By induction through the array, that is, by applying Eq.~\ref{lp_cond} to successive subdomains, we obtain that $\psi(\alpha_1 - L_1/2) = \lambda \psi(\alpha_N + L_N/2)$.
This in turn leads to {\it both} of the following conditions for the form of the $S$-matrix and for the sum of boundary phases of the LP eigenfunctions:
\begin{subequations}
 \begin{align}
  r &= \tilde{r}   \label{sac_cond} \\ e^{2ikx_c} &= \lambda,
 \end{align}
\end{subequations}
where $x_c = (\alpha_1 - L_1/2 + \alpha_{N_{\mathcal{S}}} + L_{N_{\mathcal{S}}}/2)/2$ is the array center.
This means that, for $kx_c = m\pi/2, ~m\in\mathbb{Z}$, the {\it locally} symmetric array behaves as if it were {\it globally} symmetric, with respect to its asymptotic transport properties (since $r = \tilde r$).
In particular, the invariant probability current for SAC, 
\begin{equation}
 j_k = k(1 - |t + r|^2) = -k(1 - |t + \tilde{r}|^2),
\end{equation}
vanishes with $r = \tilde{r}$ ($\neq 0$ in general). 
This can also be deduced from the unitarity condition $S^\dagger S=I$, or from the fact that $r = \tilde{r}$ yields a phase difference $\eta = \pi /2$ between $t$ and $r$, as seen from Eq.~(\ref{s_matrix}), so that $|t + r|^2 = |t|^2 + |r|^2 = 1$ and thus $j_k = 0$.

Such zero-current states connect the concept of LP to a transport observable, the current $j_k$ for SAC, in a situation where the transmission coefficient $T$ is ambiguous due to the indistinguishability of reflected and transmitted wave amplitudes in the asymptotic regions.
Nevertheless, as already seen in the previous section, definite LP is not a necessary condition for zero-current states.
Straightforward cases are perfect transmission resonances, $T = 1$ (or total reflection, $R = 1$), to be studied in the next subsection, for which the current under SAC always vanishes, without the state being LP definite.
But also in general, zero-current states exist under SAC irrespectively of LP conservation, as can be seen from Eq.~(\ref{SE_mod}) for $j_k = 0$: this linear equation for the modulus $u_k(x)$ is then fulfilled for any phase function $\varphi_k(x)$, which {\it can} yield LP, but will generally not.

\subsection{Perfect transmission resonances}

Keeping the current $j_k$ as an observable, we now turn to the case of AAC, which is most common in transport setups; 
in this case the transmission coefficient $T$ is unambiguous.
The incoming amplitudes in $|\pm \rangle$-space are now $\bigl(\begin{smallmatrix} 1 \\ 0 \end{smallmatrix}\bigr)$, that is, $\psi^<_k(x) = e^{ikx} + re^{-ikx}$ and $\psi^>_k(x) = te^{ikx}$.
Since we aim to study transmission resonances, we consider the case in which the probability current $j_{k} = kT$ does not vanish by total reflection on the potential.
A non-zero current implies the violation of Eq.~(\ref{lp_cond_phi}): if it would hold, then the spatial invariance of the current $j_{k}(x)=j_{k}(2 \alpha_n - x)$ leads to $\rho_k(x) = -\rho_k(2 \alpha_n - x) = 0$ (since $\rho_{k}$ is positive definite) in Eq.~(\ref{current}), i.e. $j_k = 0$.
Thus, (necessarily even) LP can only be fulfilled for the density $\rho_k(x)$, Eq.~(\ref{lp_cond_rho}), but not for the phase $\phi_k(x)$, in the explicitly symmetry-broken case of AAC.
This LP symmetry, left behind in the {\it modulus} of the complex field $\psi_k(x)$, can be regarded to constitute a {\it remnant} of total LP symmetry.

To see the impact of this scenario on transport, we now impose (even) LP conditions only on $\rho_k(x)$ in each array subdomain according to Eq.~(\ref{lp_cond_rho}).
This leads, again by induction through the array, to two possibilities for the overall reflection amplitude:
\begin{subequations}
  \begin{align}
   r &= 0  \label{aac_cond} \\  {\rm or}~~~~ r &= -e^{2ikx_a}
  \end{align}
\end{subequations}
where $x_a = \alpha_1 - L_1/2$ is the left end of the array, corresponding to (a) perfect transmission ($T=1$), or (b) total reflection ($R=1$).
We note that the case of total reflection \cite{tot_ref} is simply distinguished from the $T>0$ case, by the fact that the latter does not allow for nodes in $\rho_k(x)$, since then $j_k = 0$ from Eq.~(\ref{current}).
Consequently, a PTR does occur whenever $\rho_k(x)$ is completely LP symmetric within the scattering region and non-zero everywhere.
In contrast to the case of zero-current states previously discussed, here also the inverse holds:
For a PTR to occur which resonates within locally symmetric potential units, $\rho_k(x)$ {\it must} be completely LP symmetric, as proven in the Appendix.

We thus conclude the following central result of the paper:
The scattering state under AAC exhibits a PTR which resonates within locally symmetric potential units, {\it if and only if} its probability density is completely LP symmetric within the interaction region (and non-zero everywhere). 

This main result generalizes the relation between resonant transmission and symmetric probability density, from globally to locally symmetric potentials, in a conceptually transparent way.

{\it Global parity symmetry}: For a single, {\it globally} symmetric scatterer, an isolated \cite{yamamoto1995} resonance at momentum $k = k_r$ can be shown to have symmetric $\rho_k(x)$ and, therefore, be perfectly transmitting.
This is a special case of the proof given in the Appendix, but can also be derived using an expansion of the scattering state into parity definite resonant states  \cite{calderon}.
The global symmetry of $\rho_k(x)$ at a PTR simply expresses the fact that the observable, stationary  probability density of a perfectly transmitting state cannot reveal the direction of incidence in a reflection symmetric potential.

\begin{figure}[t!]
\includegraphics*[width=.9\columnwidth]{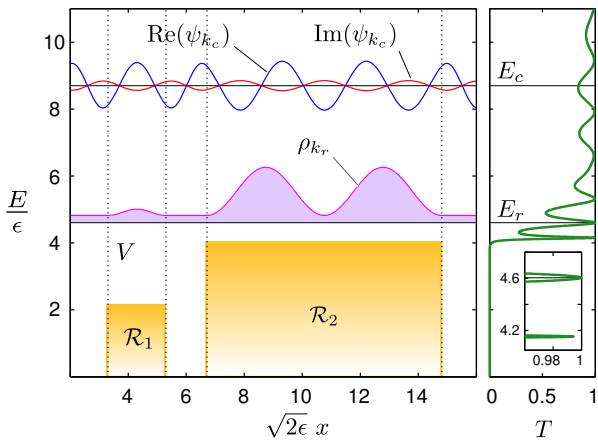}
\caption{\label{fig3} (Color online) Transmission spectrum of the simplest LP-symmetric barrier setup, with $\psi_{k_c}(x)$ and $\rho_{k_r}(x)$ shown (in arbitrary units) for a zero-current state ($E_c = 8.70\epsilon$) under SAC and a PTR ($E_r = 4.60\epsilon$) under AAC, respectively.}
\end{figure}

{\it Local parity symmetry}: The symmetry of resonant states is not directly evident, though, if different such (symmetric) scatterers are connected into a {\it locally} symmetric array.
Only when treated {\it separately} could each scatterer be argued to transmit resonantly at the same $k_r$ with corresponding globally symmetric $\rho_k(x)$ for the considered scatterer.
This would then remain unaffected by implementation into an array, since only phase-shifted plane waves propagate in the intervening potential-free regions.
However, as already anticipated from Fig.~\ref{fig1}, this is only one (the one with minimal subdomain sizes) out of many possible decompositions of the potential into LP symmetric units.
At a PTR,  not {\it all} local symmetries of the potential are necessarily followed by the resonant probability density, as will be demonstrated explicitly in Sec.~\ref{construction}.
As mentioned in Sec.~\ref{intro}, a generic LP symmetric potential possesses symmetries at different scales, with nested axes of symmetry (see Fig.~\ref{fig1}), so that the decomposition in symmetric scattering units is crucial for the identification of a PTR state.

Moreover, equivalence between separate scatterers and connected arrays thereof does not carry over to the case of zero-current states under SAC.
The proposed concept of LP treats the non-symmetric system globally while addressing its local symmetries in arbitrary subdomains, and applies uniformly to scattering under SAC or AAC.

Note that the strength of the above result for PTRs lies in its generality.  
In particular, the necessary and sufficient condition for PTRs allows us to identify all possible decompositions of the considered type of potentials (as shown schematically, e.g., for the setup in Fig.~\ref{fig1}(a)) which can support such PTRs.
As it will be demonstrated in the next section, this constitutes the basis for a construction principle for globally non-symmetric 1D scattering devices, which become resonantly transparent at prescribed energies.

\begin{figure*}[t!]
\floatbox[{\capbeside\thisfloatsetup{capbesideposition={right,top},capbesidewidth=4cm}}]{figure}[\FBwidth]
{\caption{(Color online) Potential with two resonators ${\mathcal R}_1$ (with $n_1 = 5$ barriers) and ${\mathcal R}_2$ ($n_2 = 9$), exhibiting two close (tunneling) PTRs at $E_1 = 7.00\epsilon$ and $E_2 = 7.46\epsilon$, with corresponding probability densities shown (in arbitrary units) for AAC. Their reducibility within the resonators is indicated by vertical dotted lines.}\label{fig4}}
{\includegraphics*[width=12cm]{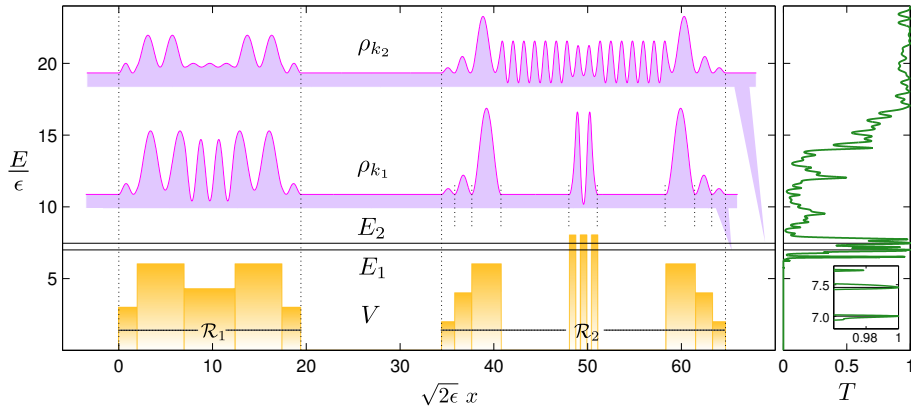}}
\end{figure*}

\section{Design of PTR states through local parity\label{construction}}

\subsection{Construction and reducibility of PTRs in \\ generic LP symmetric potentials}

Let us now proceed to the development of a construction principle of 1D devices supporting PTR states. 
This amounts to the inverse of the problem treated so far:
having shown that the PTR probability density follows the symmetries of a generic completely LP symmetric potential, we now inquire {\it how} such a potential should be designed, given that PTRs occur at desired energies.

We consider a scattering potential of the type in Eq.~(\ref{potential}), consisting of $N_{\mathcal S}$ locally symmetric units (see Fig.~\ref{fig1}(a)).
We assume, without loss of generality, that each potential function $V_n(x)$ has compact support within the $n$-th scatterer (i.e. vanishes for $|x - \alpha_n| > L_n/2$) and can be described by $N_n$ parameters $\vec{\nu}^{(n)}=(\nu^{(n)}_1, \nu^{(n)}_2,..,\nu^{(n)}_{N_n})$.
The total unimodular transfer matrix of the device, connecting the $|\pm\rangle$-amplitudes of $\psi_k^<$ to those of $\psi_k^>$, is then given by the product (ordered in $n$)
\begin{equation}
 M = \left( \begin{array}{cc} w & z \\ z^* & w^* \end{array} \right) = \prod_{n=1}^{N_{\mathcal S}} M_n(k;\vec{\nu}^{(n)}, \alpha_n),
 \label{m_matrix}
\end{equation}
with $w = (t^*)^{-1}$ and $z = -r^*(t^*)^{-1}$, where
\begin{equation}
 M_n = \left( \begin{array}{cc} w_n(k;\vec{\nu}^{(n)}, \alpha_n) & z_n(k;\vec{\nu}^{(n)}, \alpha_n) \\ z^*_n(k;\vec{\nu}^{(n)}, \alpha_n) & w^*_n(k;\vec{\nu}^{(n)}, \alpha_n) \end{array} \right)
\end{equation}
is the (unimodular) transfer matrix of the $n$-th scatterer, which we henceforth regard as known (analytically or numerically).

As discussed above, local symmetries are generally manifest on multiple scales and with different symmetry axes within the same scatterer array (see Fig.~\ref{fig1}(a)).
Therefore, the total potential can be decomposed in $N_{\mathcal D}$ {\it different} ways into LP symmetric units, each containing a number of scatterers.
In the $i$-th such decomposition, the $N_{\mathcal{S}}$ scatterers are thus grouped into $N_{\mathcal{R}}^{(i)} \leqslant N_{\mathcal{S}}$ LP symmetric {\it resonators} $\mathcal{R}_l^{(i)} ~ (l = 1,2,...,N_{\mathcal{R}}^{(i)})$ occupying the subdomains $\mathcal{D}_l^{(i)}$, separated by regions $\bar{\mathcal{D}}_{\bar{l}}^{(i)}$ ($\bar{l}=1,2,...,N_{\mathcal{R}}^{(i)}-1$) of zero potential. 
The $l$-th resonator of the $i$-th decomposition contains $n_l^{(i)}$ scatterers at positions $\{\alpha_n\}_l^{(i)}$, whose potentials are described by the set $\{\vec{\nu}^{(n)}\}_l^{(i)}$ of parameter vectors, yielding a transfer matrix $M_l^{(i)}$.

We now apply the condition $r_l^{(i)} = 0$ at $k = k_r^{(i)}$ for PTR with AAC to the scattering amplitudes through each $\mathcal{R}_l^{(i)}$ in the considered $i$-th resonator decomposition (e.g., the one depicted by the red solid line below the setup in Fig.~\ref{fig1}(a)), that is,
\begin{equation}
z_l^{(i)}(k_r^{(i)};\{\vec{\nu}^{(n)}\}_l^{(i)}, \{\alpha_n\}_l^{(i)}) = 0 
\label{z_cond_ptr}
\end{equation}
for each subdomain $\mathcal{D}_l^{(i)}~ (l = 1,2,...,N_{\mathcal{R}}^{(i)})$.

The existence of a simultaneous solution of Eqs.~(\ref{z_cond_ptr}) for all resonators $\mathcal R_l^{(i)}$ of the $i$-th decomposition, with respect to the parameters $\vec{\nu}^{(n)}$ and positions $\alpha_n$ of the scatterers, provides us with a setup with a PTR at a prescribed $k_r^{(i)}$
It is thus shown how the desired PTR can, in principle, be constructed with the aid of the derived one-to-one correspondence to LP symmetry.

Each resonator decomposition $i \in U_{\mathcal D} \equiv \{1,2,..,N_{\mathcal D}\}$ corresponds to a possible PTR at certain momentum $k_r^{(i)}$, as described above.
To construct multiple PTRs for a {\it subset} $U_{\mathcal D}^{\rm PTR} \subset U_{\mathcal D}$ [e.g., a second decomposition in Fig.~\ref{fig1}(a) is depicted by the green dashed line], the set of $\sum_{i \in U_{\mathcal D}^{\rm PTR}}N_{\mathcal R}^{(i)}$ complex equations, Eqs.~(\ref{z_cond_ptr}), must be solved {\it simultaneously for all} $i \in U_{\mathcal D}^{\rm PTR}$.
These equations determine an equal number of suitably chosen parameters among the total parameters $\vec{\nu}^{(n)}, \alpha_n ~(n = 1,2,..,N_{\mathcal S})$ of the potential.
To the remaining potential parameters, and to the desired PTR momenta $k_r^{(i)}$, fixed values are assigned in Eqs.~(\ref{z_cond_ptr}).
Note that the existence of different resonator decompositions requires, in general, multiple scatterers to be identical (e.g., in Fig.~\ref{fig1}(a) there are only three types of scatterers), which thereby reduces the number of {\it different} parameters in the $\vec{\nu}^{(n)}$.
The choice of fixed versus determined parameters depends on the specific type of considered potential.

From the described procedure, we see how LP is used to decrease the (typically vast) space of possible configurations of the considered type of potential, Eq.~(\ref{potential}), for the construction of PTRs at given energies.
The same procedure can also be used to construct zero-current states with definite LP, in which case though, the LP conditions, Eq.~(\ref{lp_cond}), must be imposed also in the gap subdomains $\bar{\mathcal D}_{\bar l}$ between the resonators.
While the restriction of parameter space through LP is achieved on generic symmetry grounds, the actual solution of Eqs.~(\ref{z_cond_ptr}) is potential specific.
In the following subsection, the construction principle will be applied to the case of PWC potentials, where the transfer matrices $M_l^{(i)}$ are expressed analytically in terms of the parameters $\vec{\nu}^{(n)}$.

To complete the general discussion, we note that the probability density $\rho_k(x)$ of a PTR state for a specific resonator decomposition may be LP symmetric within {\it smaller} subdomains than a resonator.
If the system resonates within such smaller constituents (scatterers or gaps) covering a resonator, then we refer to the state as {\it reducible} in that resonator.
The invariance of a given resonant energy under interchange or translation of resonator constituents implies their independence within the total system \cite{renner2007}, in terms of transmission.
In this sense also resonators are independent constituents of the scatterer array at PTRs.\\

\subsection{Piecewise constant potentials\label{PWC}}

For a PWC potential we have the additional restriction
\begin{equation}
 V_n(x) = V_n^0, ~n=1,2,...,N_{\mathcal{S}}
\end{equation}
in Eq.~(\ref{potential}), where the constant potential strength $V_n^0$, width $L_n$ and location $\alpha_n$ of the $n$-th barrier are freely adjustable.
The previously defined scatterer parameter vectors are now two-dimensional, $\vec{\nu}^{(n)} = (V_n^0, L_n)$.
In the following, we will consider the case of potential {\it barrier} arrays, that is, $V_n^0 > 0$ for all $n$.
The potential barriers themselves and the regions between them define the smallest possible subdomains with local symmetry and mutually different potential strengths.

The $N_{\mathcal{R}}$ resonators $\mathcal{R}_l$ of a LP decomposition are now of three types with respect to the PTRs they support: 
(a) a single barrier, supporting above-barrier resonances (ABRs), 
(b) a homogeneous array of identical equidistant barriers, supporting tunneling resonances and ABRs, or 
(c) an inhomogeneous barrier array, supporting isolated PTRs (tunneling or ABR) for appropriate combinations of barrier strengths and widths \cite{yamamoto1995}.

The $|\pm\rangle$-amplitudes along the array are determined by matching the wave function at all scatterer interfaces, and thereby connected by analytically determined single-barrier transfer matrices $M_n(k;V_n^0,L_n,\alpha_n)$.
The total transfer matrix of the device is given in analogy with Eq.~(\ref{m_matrix}), and the LP symmetry conditions applied at each interface of a given resonator decomposition $i$ lead to Eqs.~(\ref{z_cond_ptr}) with $z_l^{(i)} = z_l^{(i)}(k^{(i)};\{V_n^0,L_n\}_l^{(i)},\{\alpha_n\}_l^{(i)})$.

\begin{figure*}[t!]
\floatbox[{\capbeside\thisfloatsetup{capbesideposition={right,top},capbesidewidth=4cm}}]{figure}[\FBwidth]
{\caption{(Color online) Potential decomposed into $N_{\mathcal R} = 5$ resonators, exhibiting a zero-current state (with shown ${\rm Re}(\psi_{k_r})$ for SAC and $\rho_{k_r}$ for AAC, in arbitrary units) at energy $E_r = k^2_r/2 = 7.354\epsilon$. If ${\mathcal R}_1$ is removed (dashed lines), the state becomes a PTR (with shown $\rho_{k_r}$ for AAC, in arbitrary units) at the same energy.}\label{fig5}}
{\includegraphics*[width=12cm]{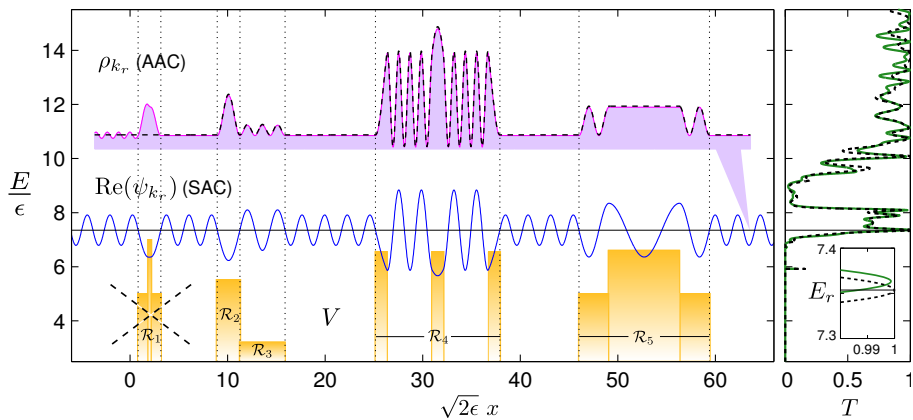}}
\end{figure*}

\begin{figure*}[t!]
\floatbox[{\capbeside\thisfloatsetup{capbesideposition={right,top},capbesidewidth=4cm}}]{figure}[\FBwidth]
{\caption{(Color online) Finite array of $N_{\mathcal S} = 12$ rectangular barriers, two of which are deformed (defects). The array is decomposed into $N_{\mathcal R}^{(1)} = 5$ and $N_{\mathcal R}^{(2)} = 12$ resonators with respect to supported irreducible PTRs at energies $E_1 = 5.00\epsilon$ and $E_2 = 8.47\epsilon$ ($\rho_{k_1}$ and $\rho_{k_2}$ are plotted in arbitrary units).}\label{fig6}}
{\includegraphics*[width=12cm]{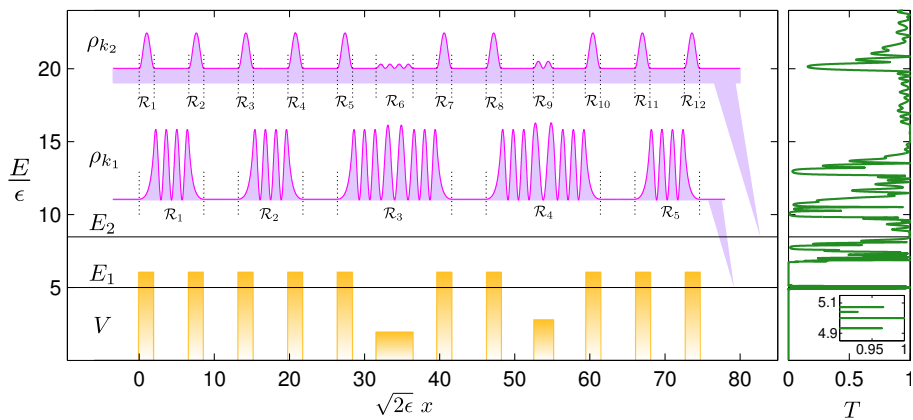}}
\end{figure*}

Let us now discuss some examples which illustrate the above concepts and their implementation more explicitly, in order of increasing resonator number $N_{\mathcal R}$.
Fig.~\ref{fig3} demonstrates the occurrence of a zero-current state and an irreducible PTR in the simplest case which breaks global parity, an asymmetric double-barrier setup.
As we see, the zero-current state has even LP in the barriers and odd LP between them.
$T(E)$ displays a multi-ABR structure, but only the depicted peak corresponds to a PTR state (see the inset).

Fig.~\ref{fig4} shows again a setup of $N_\mathcal{R} = 2$ resonators, but of a more complex structure, supporting two energetically close PTRs.
$\rho_{k_2}$ (at $E_2$) is irreducible in both resonators, while $\rho_{k_1}$ (at $E_1$) is irreducible in ${\mathcal R}_1$ but reducible in ${\mathcal R}_2$ (as shown by the vertical dotted lines), where it varies only within the barrier regions, while flat $\rho(x)$ indicates a forward propagating wave only.
This is an example of LP being fulfilled at two different scales (here within resonator ${\mathcal R}_2$) in the same system. 
Note that the occurrence of the PTR at $E_{2}$ requires that the positioning of scattering units in ${\mathcal R}_2$ supports LP symmetry at a new spatial scale, in accordance with the general construction principle described above.
We also point out that the setup remains resonant for arbitrary combinations of $\mathcal{R}_1$ and $\mathcal{R}_2$, with repeated corresponding patterns in $\rho_k(x)$.
A system with the resulting structural complexity can thus support PTRs, relying on (repeated) LP symmetries in the potential.

In Fig.~\ref{fig5} a zero-current state becomes a PTR at the same energy by removing the first resonator ${\mathcal R}_1$ from an $N_{\mathcal{R}}=5$ array.
As can be seen, the PTR state is overall irreducible, and more localized within ${\mathcal R}_4$.
Notice that this PTR is independent of the width of the central barrier in ${\mathcal R}_5$, within which the wave propagates only forwardly with constant $\rho_{k_r}(x) > 1$.
This might be utilized for the flexible design of efficiently transmitting non-symmetric devices.

In Fig.~\ref{fig6} we investigate a finite periodic array with two defects in the form of alternate lattice cells.
Without the defects, the transmission properties of the uniform array are determined by its unit cell \cite{griffiths2001}, which also defines the scale of local symmetry.
Due to coupling of the degenerate resonant levels of adjacent identical resonators, a uniform $N$-scatterer array exhibits $(N-1)$-fold split PTRs \cite{griffiths2001, morfonios2009}, which saturate into transmission bands for increasing N.
As we see, the presence of aperiodicity~\cite{hernandez2010} distorts the precursors of the energy bands~\cite{luna2009,Dietz2012} and lowers the resonances in $T(E)$ from unity because of the induced asymmetry~\cite{Ferry1997}.
Nevertheless, the decomposition into resonators [particularly of type (c)] containing multiple barriers reveals the possibility of PTRs, as explained above, owing to the locally symmetric $\rho_{k_r}$ of irreducible resonant states.

This implies that identification of local symmetries on scales larger than the minimal constituent building blocks provides a key for the description of structurally complex systems.
Moreover, different LP axes (i.e. resonator decompositions) for the same setup correspond to different PTR levels;
local symmetry considerations thus prove to be of fundamental importance in accessing and understanding the properties of aperiodic non-uniform systems.

\section{Conclusion \label{conclusion}}

We have introduced the concept of local parity (LP) and revealed its impact on the transport properties of aperiodic 1D arrays of arbitrary reflection symmetric scatterers. 
The manifestation of LP for the potential and for a generic quantum state was simply reduced to the domain-wise invariance or vanishing of a derived non-local quantity.
Scattering states were generically classified with respect to their LP properties with the aid of this invariant quantity as well as the probability current.
It was shown (i) that eigenstates of total LP operators carry zero current for symmetric asymptotic conditions, and (ii) that a remnant of LP symmetry in the wave-function modulus underlies the emergence of perfect transmission resonances (PTRs) which resonate within locally symmetric potential units.
Consequently, the decomposition of globally non-symmetric arrays into different symmetric resonator units relates perfect transmission to LP symmetries of the stationary probability density.
PTR states were shown to depend on their spatial reducibility into LP symmetric parts of altering sizes and symmetry axes arrangements, even within the same potential.
This in turn demonstrates the importance of considering local order on different scales to understand the behavior of systems with structural complexity.
Invariance of resonant transmission under translation or exchange of resonator subdomains also links to the concept of independence among constituents of extended systems.
Our findings were demonstrated by applying a general construction principle for PTRs to the analytically solvable case of piecewise constant potentials.
The generalization of our approach to higher dimensions and different kinds of local symmetry transformations could provide a different context for the analysis of complex systems, based on fundamental principles.

\section{Acknowledgments}

The authors are thankful to an anonymous referee, whose valuable remarks helped to improve the manuscript.
This research has been co-financed by the  European Union (European Social Fund - ESF) and Greek national funds through the Operational Program ``Education and Lifelong Learning" of the National Strategic Reference Framework (NSRF) - Research Funding Program: Heracleitus II. Investing in knowledge society through the European Social Fund.\\

\begin{center}
{\bf APPENDIX }
\end{center}

We prove here that a PTR which resonates within locally symmetric potential units, occurs at an energy $E = k^2/2$ {\it only if} its probability density $\rho_k(x)$ is completely LP symmetric.
This is equivalent to stating that, if such a PTR occurs at $E$, then $\rho_k(x)$ is completely LP symmetric.

\begin{figure}[t!]
\includegraphics*[width=.8\columnwidth]{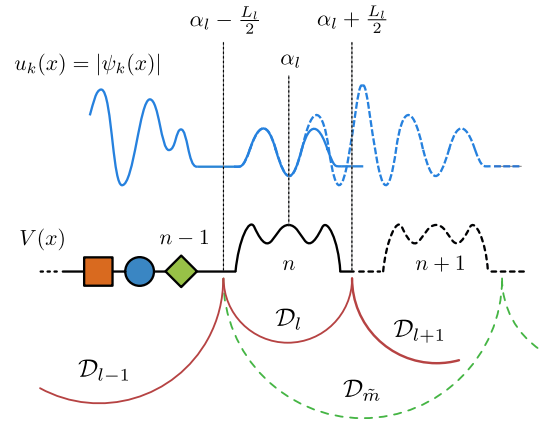}
\vspace{.2cm}
\caption{\label{proof} (Color online) Part of the potential $V(x)$ of around the $n$-th scatterer, with corresponding wave function modulus $u_k(x)$. Either (i) a subdomain ${\mathcal D}_l$ is itself perfectly transmitting with LP symmetric $u_k(x)$ (solid lines), which is shown by considering a BVP in each of its halves (see the Appendix), or (ii) it must be augmented by a subsequent subdomain, so that $u_k(x)$ is LP symmetric in the resulting subdomain ${\mathcal D}_{\tilde m}$ (dashed lines).}
\end{figure}

First, we prove that, if the transmission $T_l$ through a single LP symmetric subdomain ${\mathcal D}_l$ is perfect, $T_l = 1$, then $\rho_k(x)$ is LP symmetric within ${\mathcal D}_l$ (for a schematic illustration, see Fig.~\ref{proof}), as follows.

If $T_l = 1$, then at the boundaries of ${\mathcal D}_l$ the conditions $u_k(x=\alpha_l-L_l/2) = u_k(x=\alpha_l+L_l/2) = T_l = 1$ and $u'_k(x=\alpha_l-L_l/2) = u'_k(x=\alpha_l+L_l/2) = 0$ apply. 
In the left half ${\mathcal D}_l^L \equiv [\alpha_l-L_l/2, \alpha_l]$ of ${\mathcal D}_l$, the unique solution $\psi_k(x)$ of Eq.~(\ref{SE_psi}) under AAC has modulus $u_{k,L}(x)$ which obeys the boundary value problem (BVP) 
\begin{subequations}
  \begin{align}
 \frac{1}{2} u_{k,L}''(x) - \frac{j_k^2}{2u_{k,L}^3(x)} + \gamma^2(x)u_{k,L}(x) &= 0, ~ \\ x &\in {\mathcal D}_l^L \nonumber \\
  u_{k,L}(x)\rvert_{x = \alpha_l-L_l/2} &= 1, \\
  u'_{k,L}(x)\rvert_{x = \alpha_l-L_l/2} &= 0, \\
  u_{k,L}(x)\rvert_{x = \alpha_l} &= u_k^{\alpha_l},
  \end{align}
\label{bvp_l}
\end{subequations}\\
where $u_k^{\alpha_l} \equiv u_k(\alpha_l) = |\psi_k(\alpha_l)|$ \cite{uniqueness}.
In the right half ${\mathcal D}_l^R \equiv [\alpha_l, \alpha_l + L_l/2]$ of ${\mathcal D}_l$, the modulus $u_{k,R}(x)$ obeys the BVP
\begin{subequations}
  \begin{align}
 \frac{1}{2} u_{k,R}''(x) - \frac{j_k^2}{2u_{k,R}^3(x)} + \gamma^2(x)u_{k,R}(x) &= 0, ~ \\ x &\in {\mathcal D}_l^R \nonumber \\
  u_{k,R}(x)\rvert_{x = \alpha_l + L_l/2} &= 1, \\
  u'_{k,R}(x)\rvert_{x = \alpha_l + L_l/2} &= 0, \\
  u_{k,R}(x)\rvert_{x = \alpha_l} &= u_k^{\alpha_l}.
  \end{align}
\label{bvp_r}
\end{subequations}

Under a {\it passive} transformation $x \to 2\alpha_l - x$ of only the reference coordinate system (that is, keeping the potential intact), and using the LP symmetry $\gamma(x) = \gamma(2\alpha_l - x)$ for $x \in {\mathcal D}_l$, the {\it same} BVP for $u_{k,R}(x)$, Eq.~(\ref{bvp_r}), reads
\begin{subequations}
  \begin{align}
 \frac{1}{2} u_{k,R}''(2\alpha_l - x) - \frac{j_k^2}{2u_{k,R}^3(2\alpha_l - x)} ~~~&\\ + \gamma^2(x)u_{k,R}(2\alpha_l - x) = 0, ~~ x &\in {\mathcal D}_l^L \nonumber \\
  u_{k,R}(2\alpha_l - x)\rvert_{x = \alpha_l - L_l/2} &= 1, \\
  u'_{k,R}(2\alpha_l - x)\rvert_{x = \alpha_l - L_l/2} &= 0, \\
  u_{k,R}(2\alpha_l - x)\rvert_{x = \alpha_l} &= u_k^{\alpha_l}.
  \end{align}
\label{bvp_r_transf}
\end{subequations}
Comparison of the BVPs in Eqs.~(\ref{bvp_l}) and (\ref{bvp_r_transf}) for the functions $u_{k,L}(x)$ and $u_{k,R}(2\alpha_l - x)$ of $x$ yields
\begin{equation}
 u_{k,L}(x) = u_{k,R}(2\alpha_l - x),
\end{equation}
which shows that $u_k(x)$ is necessarily LP symmetric within ${\mathcal D}_l$.

Having proven that $T_l = 1$ leads to LP symmetric $\rho_k(x)$ in ${\mathcal D}_l$, we now {\it assume} that the transmission through the whole array is perfect, $T = 1$, for a state which resonates within locally symmetric potential units, and we show that then $\rho_k(x)$ is {\it completely} LP symmetric.
To do this, we start at the left end $x = \alpha_1 - L_1/2$ of the potential $V(x)$ of Eq.~(\ref{potential}), and consider the {\it first smallest} LP symmetric subdomain ${\mathcal D}_l$ ($n = 1$) with non-zero potential.
Now, depending on whether the case (i) $T_l = 1$ or (ii) $T_m < 1$ is fulfilled for $l = m = 1$, we proceed as dictated below:

(i) $T_l = 1$: Then, $\rho_k(x)$ is LP symmetric in ${\mathcal D}_l$ (see solid lines in Fig.~\ref{proof}).
Starting from the {\it upper} boundary of ${\mathcal D}_l$, $x = \alpha_l + L_l/2$, consider its {\it subsequent smallest possible} LP symmetric subdomain with non-zero potential, ${\mathcal D}_{l+1}$.
If $T_{l+1} = 1$, repeat the present step (i) with $l \to l+1$; otherwise, if $T_{l+1} < 1$, apply step (ii) with $m \to l+1$.

(ii) $T_m < 1$: Then, $\rho_k(x)$ is not LP symmetric in ${\mathcal D}_m$ (see dashed lines in Fig.~\ref{proof}).
Starting from the {\it lower} boundary of ${\mathcal D}_m$, $x = \alpha_m - L_m/2$, consider the {\it first smallest possible} LP symmetric subdomain with non-zero potential, denoted ${\mathcal D}_{\tilde m}$, which is {\it larger} than ${\mathcal D}_m$.
If $T_{\tilde m} = 1$, proceed to step (i) with $l \to \tilde m$; otherwise, if $T_{\tilde m} < 1$, repeat this step (ii) with $m \to \tilde m$.

In this way, starting from $l = m = 1$, steps (i) and (ii) lead us through the scatterer array until the {\it last} smallest possible LP symmetric subdomain with non-zero potential is reached, either through step (i) or through step (ii).
For this last subdomain, case (i) must necessarily hold, since in total $T = 1$ by assumption, and so $\rho_k(x)$ is completely LP symmetric.
This completes the proof.


\begin{thebibliography}{99}
\bibitem{tot_ref} Although the case of $R = 1$ is equivalent in terms of LP, it breaks the symmetry of $\rho_k(x)$ even in the trivial case of a globally symmetric potential.
\bibitem{uniqueness} In general, the uniqueness of the solution of the non-linear BVP for arbitrary $\gamma(x)$ is by no means trivial. Nevertheless, it is here ensured by the fact that $u_k(x)$ is the modulus of the wavefunction $\psi_k(x)$, which we know is unique under the imposed AAC.
\end{thebibliography}
\end{document}